\input{aipcheck}
\documentclass[
    ,final            
  ]
  {aipproc}

\layoutstyle{6x9}
\newcommand{\Ub}{$\mathbf U$-boson}

\newcommand{\PR}{{\em Phys. Rev. }}
\newcommand{\PRL}{{\em Phys. Rev. Lett. }}
\newcommand{\PL}{{\em Phys. Lett. }}
\newcommand{\NP}{{\em Nucl. Phys. }}

\newcommand{\etal}{{\em et al}}
\begin{document}

\title{\bf Searching for a \Ub{} with a positron beam}


\author{B.~Wojtsekhowski}
{address={Thomas Jefferson National Accelerator Facility, 
Newport News, VA 23606, USA }}

\begin{abstract}
A high sensitivity search for a light \Ub{} by means of a positron beam 
incident on a hydrogen target is proposed. 
We described a concept of the experiment and two possible realizations.
The projected result of this experiment corresponds to an upper limit 
on the square of coupling constant $ |f_{_{eU}}|^2 = 3 \times 10^{-9}$
with a signal to noise ratio of five.
\end{abstract}

\maketitle
\vspace{-0.50in}
\subsection{Introduction}
\label{sec:introduction}
\vspace{-0.10in}
The search for an experimental signature of super symmetry, 
proposed in the mid 1970s, see e.g.~\cite{fa77}, 
is a major effort of modern particle physics, see review~\cite{pdg4}.
Most of the search activity is focused on possible heavy particles with 
a mass scale of 1~TeV and above.
At the same time, as was suggested by P.~Fayet, there could be another 
interesting $U(1)$ symmetry, which requires a new gauge boson~\cite{fa80}.
The boson could be light and weakly interacting with known particles.
Most constraints for the light \Ub{} parameters were obtained from
electron/muon $g-2$ and particle decay modes \cite{fa06, fa08}.
The possible connection of the \Ub{} and dark matter in view of the
observed positron abundance was investigated for several years 
and is often referred to as a light dark matter (LDM)\cite{bo04,fa04,bofa,bo06}.

Several methods were used in search of the \Ub{} signal.
They consider ``invisible'' decay modes of the \Ub{} and therefore result 
in an upper limit for $\mathbf {f_{_{qU}} \times B_{U \rightarrow invisible}}$.
The first method uses precision experimental data on exotic
decay modes of elementary particles, e.g. 
$\pi^\circ \rightarrow invisible + \gamma$,
for the calculation of the upper limit of the \Ub{} coupling constant 
to the specific flavor.
These upper limits for decay of the $J/\Psi$ and $\Upsilon$ to a photon
plus invisible particles were obtained experimentally
by means of the ``missing particle'' approach, where 
a missing particle in the event type $e^+e^- \rightarrow \gamma X$
leads to a yield of events with a large energy photon detected at 
a large angle with respect to the direction of positron 
and electron beams.
From the yield of such events the coupling constant could be determined
for a wide range of mass of the hypothetical \Ub.
A recent experiment~\cite{BABAR} using statistics of 
$1.2 \,\, 10^8 \,\,\, \Upsilon(3S)$ events provided the best
data for $\Upsilon(3S)$ decay to $U + \gamma$ and a 
limit on the coupling of the \Ub{} to the $b$-quark.
In the mas region of below 100 ~MeV, the limit for 
$B(\Upsilon(3S) \rightarrow \gamma A^0) \times B(A^0 \rightarrow 
\mathrm invisible)$ is $3 \cdot 10^{-6}$,
from which the limit was obtained 
${\mathbf {f_{_{bA}} < 4 \cdot 10^{-7} m_{_U}\,[\mathrm {MeV}]}}$~\cite{fa08}.
An additional hypothesis of coupling constants universality is required 
to get a bound on ${\mathbf {f_{_{eU}}}}$, so direct measurement 
of the coupling to an electron is of large interest.
Presently the upper limit on the vector coupling 
obtained from the measurement of the electron ($g-2$) is 
$\mathbf {f_{_{eU}} < 1.3 \cdot 10^{-4} \, 
m_{_U}\,[\mathrm {MeV}]}$~\cite{fa06}.

A direct measurement of $f_{_{eU}}$ could be done by detecting 
the decay of the \Ub{} to an electron-positron pair and reconstructing 
the $e^+e^-$ invariant mass.
This method also provides the particle mass.
Another complication of the invariant mass method is the high level of 
electromagnetic background in the mass spectrum of $e^+e^-$, 
so such a measurement requires very large statistics.
Several groups recently advocated the use of data sets accumulated in
collider experiments for such an analysis~\cite{bo06}.
There is also a suggestion~\cite{fi09} to use low-energy 
electron-proton scattering for the generation of the \Ub{} and 
its observation via an invariant mass of the final $e^+e^-$ pair.

A sensitive \Ub{} search could be performed with a low energy 
$e^+e^-$ collider, where several search techniques could be used:
\begin{itemize}
\item Invisible particle method
\item Invariant mass of the final $e^+e^-$ pair
\item Missing mass with single-arm photon detection. 
\end{itemize}

The electron-positron collider AdA, a pioneering storage ring
~\cite{AdA}, operated at just 100-200~MeV energy per beam.
However, to search for the \Ub{}, with its mass of 10-20~MeV, 
the center mass energy of $e^+e^-$, $E_{cm}$, should be low.
When $E_{cm}$ is below pion threshold the obtained result will be
directly related to $f_{_{eU}}$.
The production cross section is proportional to $1/E_{cm}^2$, so for 
low $E_{cm}$ even a modest luminosity would be sufficient for 
a precision measurement.  

Interestingly, low $E_{cm}$ can also be achieved without the use 
of low-energy beams,
but rather with a sliding geometry arrangement.
Sliding geometry for the beam-beam experiment means 
that the angle of collision, $\theta_{+-}$ is 
not equal to 180$^\circ$ as in all existing colliders, but is small. 
So, the value of $E_{cm}^2 = E_+ E_- \, (1 - \cos\theta_{+-})$ could be 
as low as 20~MeV with 500~MeV beams.
Such an approach allows for a compact transverse size of the beams and 
high luminosity.
Development of the ``sliding'' beam facility is a relatively large project, 
so for now we will concentrate on a much simpler proposal.
\vspace{-0.25in}

\subsection{The concept}
\vspace{-0.10in}
We propose to arrange low-energy $e^+e^-$ collisions using of
a positron beam with energy, $E_+$, of a few hundred MeV and 
a stationary hydrogen target (this proposal was reported in Ref.~\cite{wo06}).
In such a ``collider'' the energy is sufficient to search for a 
light \Ub{} with a mass up to $m_{_U} \approx \sqrt{E_+[\mathrm {MeV}]}$.
The experiment could be organized with an internal target in a
storage ring with a high intensity positron beam or 
with a positron beam incident on an external hydrogen target.
Depending on the experimental luminosity different search methods will be used.
At the highest luminosity use of the ``invisible particle'' method and 
the invariant mass of the final $e^+e^-$ method are limited
by the accidental rate.
So, the reaction of interest will be $e^+e^- \rightarrow \gamma X$ 
for which the yield is proportional 
to the second power of coupling constant, $f_{_{eU}}$.

\begin{figure}[htb]
       \includegraphics[width=0.8\textwidth]{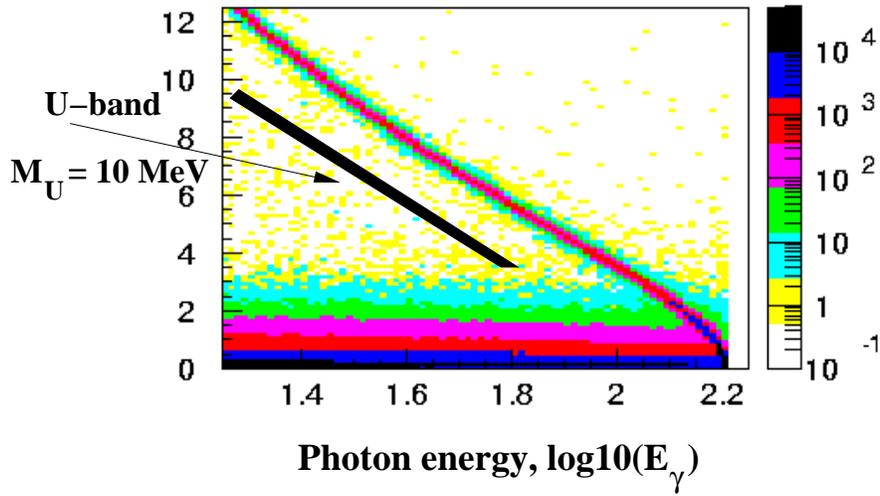}
       \caption{Two-dimensional distribution of the photon events
in the scattering angle (degrees) and the photon energy 
($log_{10}E[\mathrm{MeV}]$) for a 160~MeV positron beam
incident on a 1~cm liquid hydrogen target.
The black line presents the location of \Ub{} events of 10~MeV mass.} 
       \label{fig:gamma-2d}
\end{figure}

In the process $e^+e^- \rightarrow U \gamma$ measurement of the photon 
energy and its angle allows a reconstruction of the missing-mass spectrum 
and a search for a peak corresponding to the \Ub{}.
In such a spectrum the dominant signal corresponds
to the annihilation reaction $e^+e^- \rightarrow \gamma \gamma$.
The signal for the \Ub{} will be shifted to the area of the continuum
(see illustration in Fig.~\ref{fig:gamma-2d}).
The continuum part of the event distribution has several contributions. 
The first is the three-photon annihilation process,
the second is photon emission in positron 
scattering from an electron or a proton in the target, 
and the third, important only for a large external target thickness, 
is bremsstrahlung by the positrons passing through the target.
\vspace{-0.25in}

\subsection{\bf The kinematics and cross section}
\begin{figure}[htbp]
\begin{minipage}[b]{0.50\linewidth}
\begin{center}
        \includegraphics[width=0.75\textwidth]{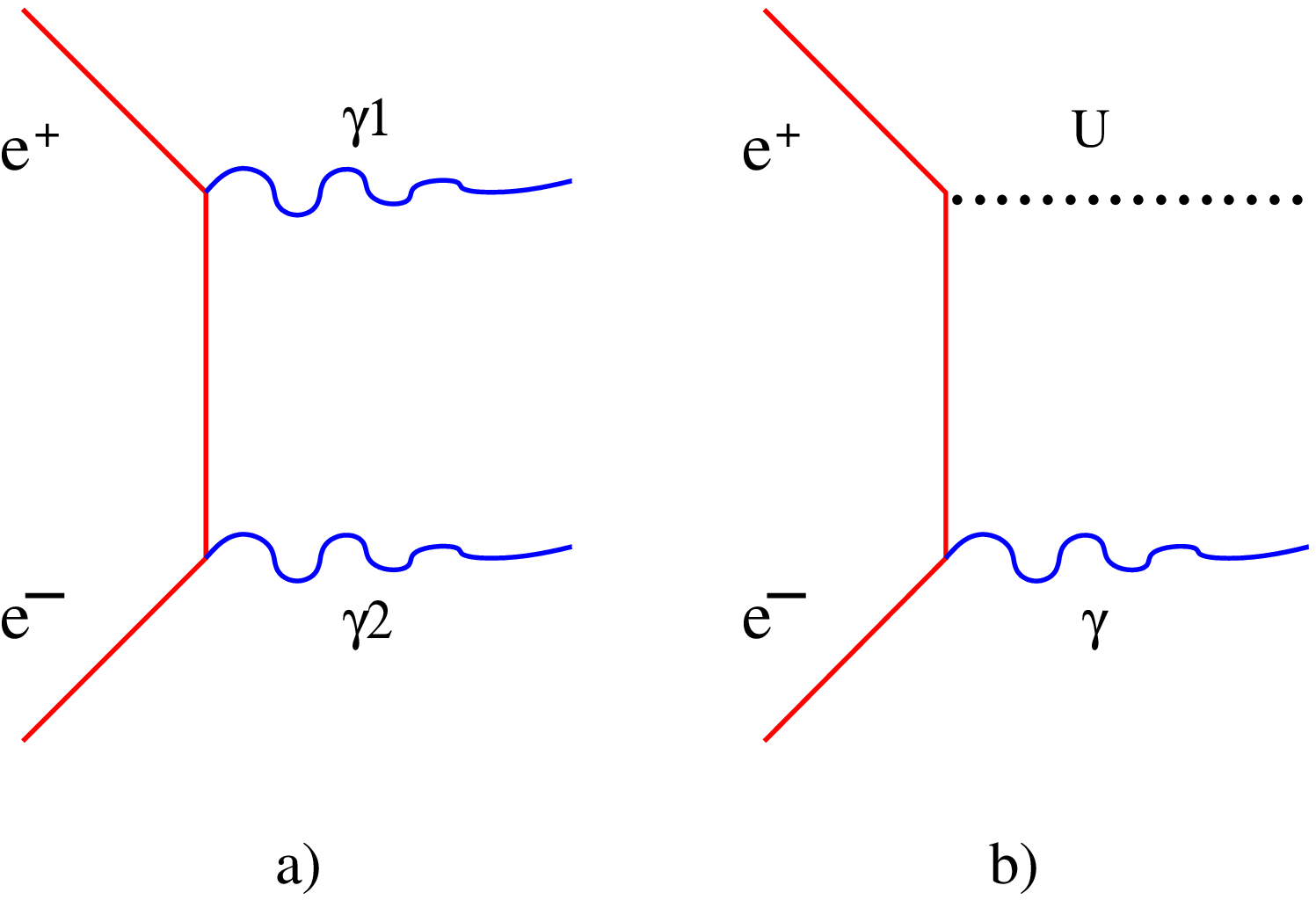}
\end{center}
\end{minipage}
\hfill
\begin{minipage}[b]{0.50\linewidth}
\begin{center}
     \includegraphics[width=0.65\textwidth]{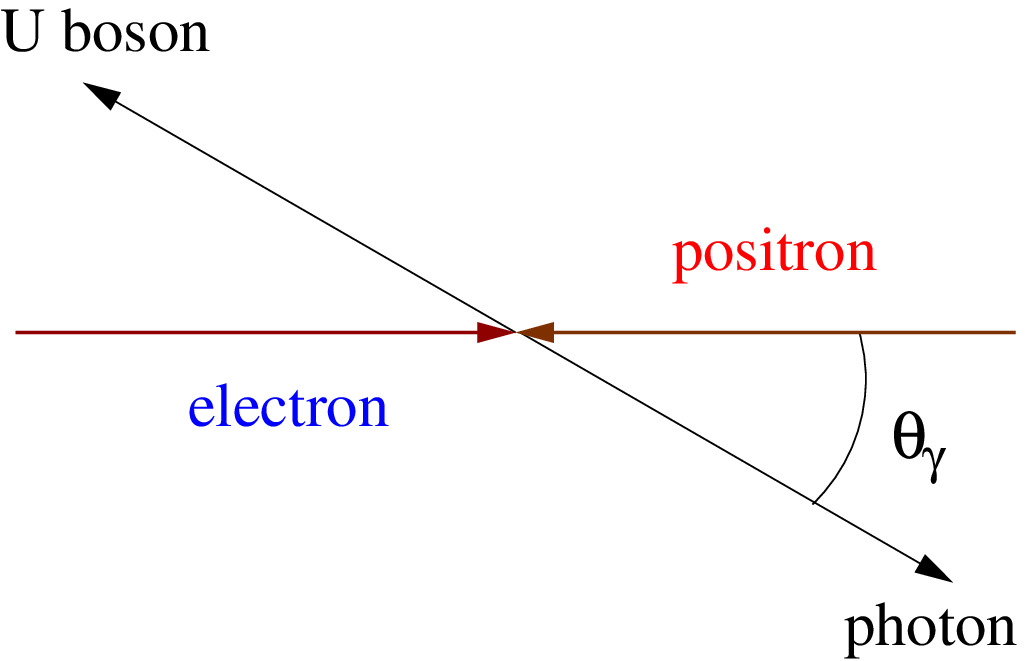}
\end{center}
\end{minipage}
\label{fig:two-body}
\caption{
The diagrams: a) two-photon annihilation, b) the \Ub$-\gamma$ production.
The $e^+e^- \rightarrow U + \gamma$ reaction.}
\end{figure}

\vspace{-0.05in}
Two-photon annihilation is a dominant process of high-energy photon 
production in $e^+e^-$ collisions at a cms energy of a few tens of MeV.
Two reactions, depicted in the left panel of Fig.~\ref{fig:two-body}, 
are two-photon annihilation and production of an exotic \Ub{}. 
The kinematics for the two-body final state is shown in the right panel
of Fig.~\ref{fig:two-body}.
The energy in the center of mass system 
$\sqrt{s} \,=\,  \sqrt{2m^2 \,+\, 2E_+ m} $,
where $m$ is the electron mass and $E_+$ the positron energy,  
and the emission angle of the final photon $\theta_\gamma$ 
with respect to the direction of positron beam
define the value of the photon energy $E_\gamma$.
In the case of two-photon production: 
$E_{\gamma (\gamma\gamma)}^{lab} \approx E_+ (1 \,-\, \cos\theta_\gamma^{cm})/2$.
In the case of \Ub{} production: 
$E_{\gamma(U\gamma)}^{lab} \,=\, E_{\gamma (\gamma\gamma)}^{lab}
\cdot (1 \,-\, M_{_U}^2/s)$.
The kinematic boost from the cm system to the lab leads to a larger photon
energy in the forward direction, which helps the measurement of 
the photon energy.
The large variation of the photon energy with the photon angle 
in the lab system provides an important handle on the systematics. 
The cross section of $e^+e^-$ two-photon annihilation
in flight was first obtained by Dirac.
In the practical case of {${\gamma_+ >> 1}$ } the total cross section is: 
{${\sigma \,=\, \frac{\pi r_e^2}{\gamma_+} \cdot (\ln 2 \gamma_+  -1)} $}.
The differential cross section of the process is:
{${\frac{d\sigma}{d\cos\theta_\gamma^{cm}} \,=\, 
\frac{2\pi r_e^2}{\gamma_+}  
\cdot \left [ \frac{1}{\sin^2\theta_\gamma^{cm}} - \frac{1}{2} \right ] }$ }.
The cross section of the \Ub{} production in $e^+e^-$ collusion 
{$(e^+e^- \rightarrow U\gamma)$} is less than the 
{$\gamma\gamma$} production cross section by the double 
ratio of the squared coupling constants, 
${\mathbf {2 \,|f_{_{eU}}|^2/e^2}}$ \cite{fa06}.
\vspace{-0.25in}

\subsection{\bf The proposed experimental setup}
\vspace{-0.10in}
For this version of the measurement we have developed, in some detail, 
a single-arm approach for the detection of the \Ub{} signal.
At the same time, we want to stress, that the ``invisible particle''
method may also provide high sensitivity and will likely need
a lower positron beam intensity, but it requires a 
more complicated detector.

\begin{figure}[htb]
        \includegraphics[width=0.85\textwidth]{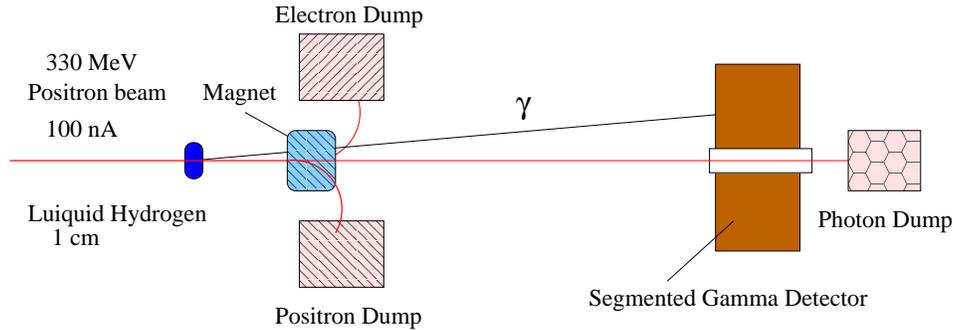}
\caption{The layout of proposed experiment with an external target.} 
\label{fig:layout1}
\end{figure}

The measurement is proposed that will allow a 5 sigma observation 
(if it exists) a dark matter gauge \Ub{} in the mass interval 2-15~MeV
with a square of coupling constant, {$\mathbf {|f_{_{eU}}|^2}$}, as low as 
{$\mathbf {3 \cdot 10^{-9}}$}.
This experiment will determine the mass of the \Ub{} and 
the value of coupling constant.
The layout of the experiment is presented in Fig.~\ref{fig:layout1}.
The positron beam will pass through the short liquid hydrogen target and 
then be sweeped out by a dipole magnet to the positron dump.
The secondary electrons will be swept out by the same magnet in the opposite 
direction.
The bremsstrahlung photons, which have a typical angle with the beam 
direction of $m/E_{beam}$ will be absorbed in the photon dump.
The photons of the electron-positron annihilation with a lab angle
in the range of 2 and 6 degrees will be detected by a segmented photon detector.
Use of the sweep magnet allows removal of the charged particles and 
reduces the rate in the detector.
However, if the DAQ readout speed is sufficient, it will be
better to detect all particles and to veto events 
originating from positron-electron scattering during off-line analysis.
Such events have associated photons and present a dominant part
of the continuum background.       

\begin{figure}[htb]
        \includegraphics[width=0.75\textwidth]{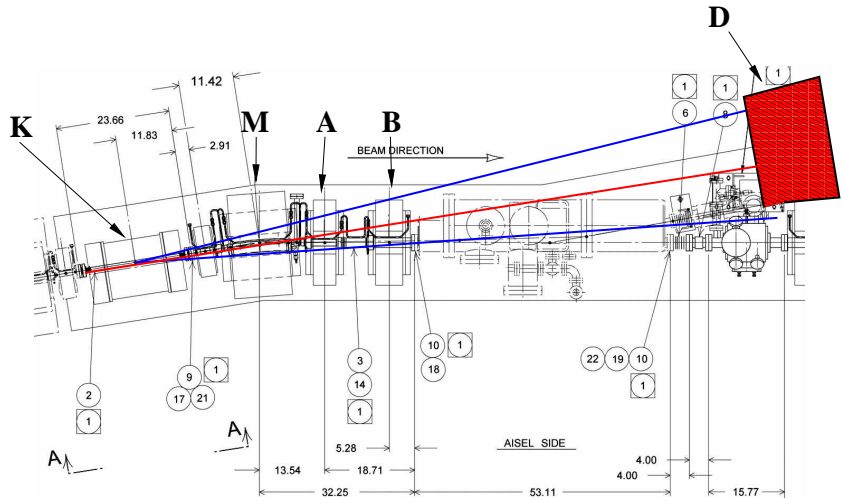}
\caption{The drawing of the extraction straight section of the SLAC
positron damping ring and the layout of proposed experiment 
with an internal target.
Red lines shows the central direction of the photons. 
Blue lines indicates $\pm 5^\circ$ acceptance to the detector {\bf D}.
Magnet {\bf M} needs a larger gap.
The two quadrupoles, ({\bf A} and {\bf B}) need to be relocated to allow
the photon path from the target to the detector.} 
\label{fig:layout2}
\end{figure}

A different arrangement of the experiment could be used with a storage ring 
as shown in Fig.~\ref{fig:layout2}.
Such an arrangement is possible at the SLAC positron damping ring and
the BINP storage ring VEPP-III.
The very small internal target thickness presents an 
important advantage in this setup.
The internal target will be placed in the straight section {\bf K}
which is followed by magnet {\bf M}.
This magnet is a part of the accelerator and is also used to sweep 
charged particles away from the photon detector {\bf D}.
\begin{figure}[htb]
\begin{minipage}[b]{0.50\linewidth}
\begin{center}
        \includegraphics[width=1.00\textwidth]{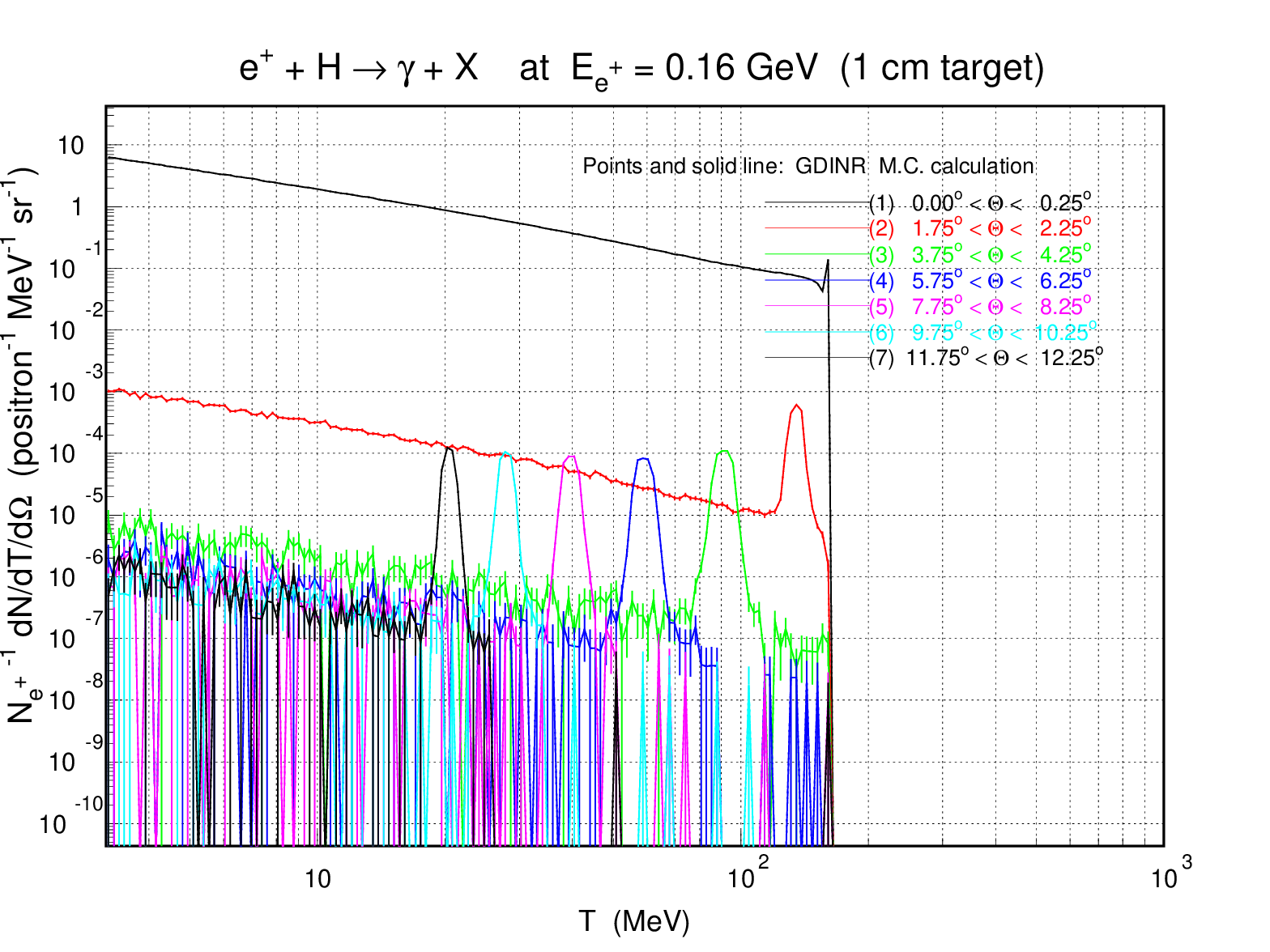}
\end{center}
\end{minipage}
\hfill
\begin{minipage}[b]{0.50\linewidth}
\begin{center}
        \includegraphics[width=1.00\textwidth]{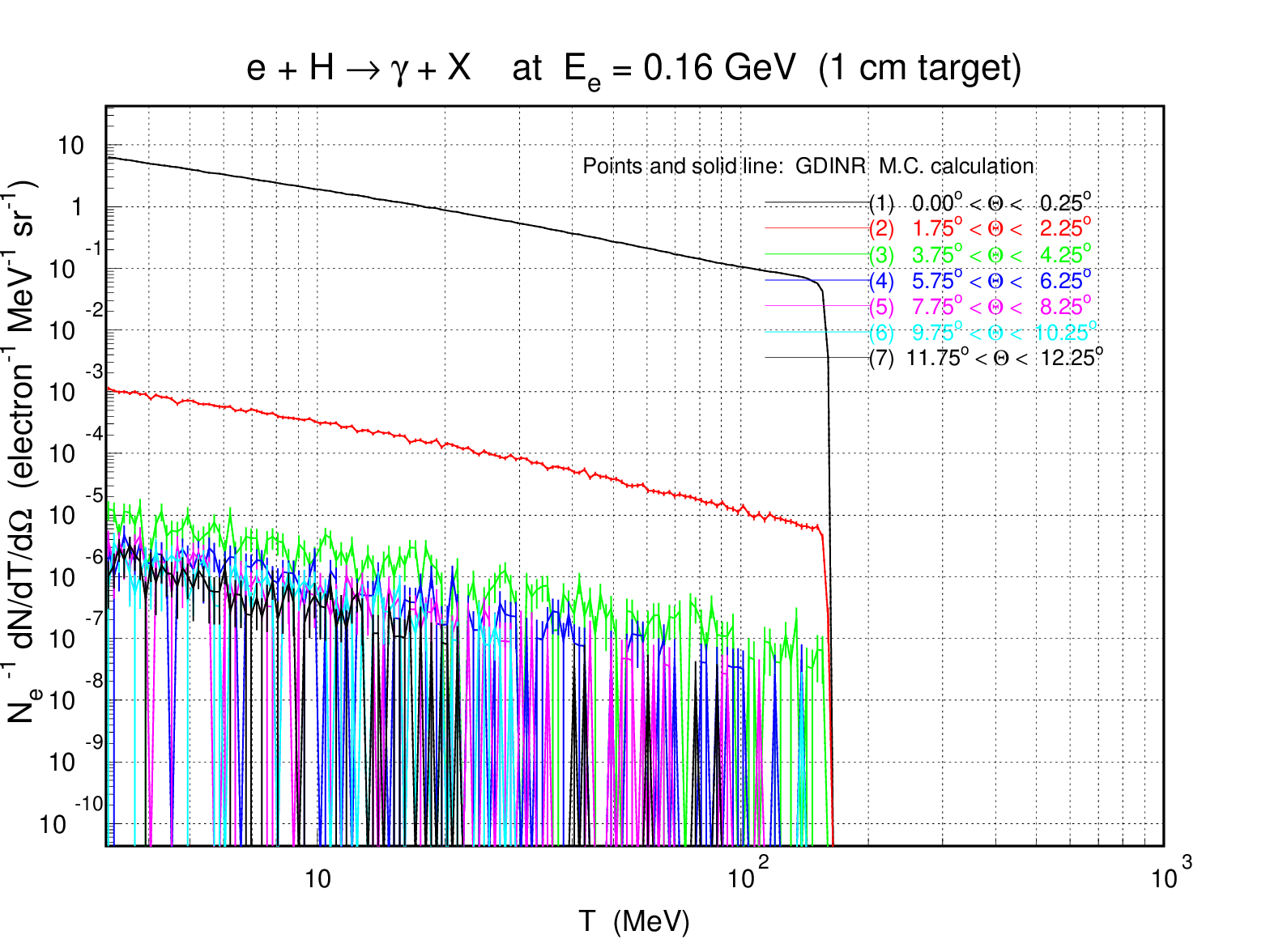}
\end{center}
\end{minipage}
\caption{The photon spectra in case of a positron beam
(left panel) and an electron beam (right panel) 
incident on 1~cm long liquid hydrogen target.}
\label{fig:MC}
\end{figure}

Fig.~\ref{fig:MC}, left panel, shows the result of a MC simulation of 
the photon spectra for the case of a 1~cm liquid hydrogen
external target and a 160~MeV positron beam energy.
The intensity of the background process in the proposed type of search 
is about 30-100 times below the main process, 
$e^+e^- \rightarrow \gamma \gamma$, whose peak
is moving with the scattering angle.
In the case of an electron beam the energy spectrum is smooth which
will be used for calibration of the detector response.

The experiment will require a careful account of the detector responses.
The energy response will be calibrated by using $\gamma-\gamma$
coincidence events produced with the liquid hydrogen target.
These data will also provide a detector line shape determination.
The photon ``white'' spectrum, created with a Be or C target will be
also used to measure the detector response density, which is a crucial 
parameter in the search for a signal, 
which could be at the level of $10^{-6}$ of the QED background.
The use of the electron beam instead of the positron beam
provides another way to obtain the photon spectra without the \Ub{} 
signal and the two-photon line. 
\vspace{-0.25in}
		   
\subsection{\bf The projected sensitivity }
\vspace{-0.10in}
The stationary target $e^+e^-$ ``collider''allows a luminosity of
at least $10^{34}$~Hz/cm$^2$ or even higher.
Considering scattering angles near $\theta_{cm} = 90^\circ$,
a 330~MeV positron beam and positron-electron 
luminosity $10^{34}$~Hz/cm$^2$,
the photon rate from the annihilation process will be
10~MHz in a $\theta_{cm}$ interval
from 60$^\circ$ to 120$^\circ$, which corresponds
to $\theta_{lab}$ from 1.9$^\circ$ to 5.5$^\circ$ .
In a three-month run the total accumulated statistics will 
be $0.8\cdot10^{14}$ events.
The continuum background at $\theta_{cm}=90^\circ$ is mainly
due to photon radiation along the final direction by
a positron and an electron in Bhabha scattering.
Assuming the relative energy resolution of the photon detector 
to be about 5\% we will use 15\%-wide energy bin for the search window.
The background rate in a 15\% photon energy interval was
estimated to be 1.2\% of the annihilation photon rate.
So, the statistical uncertainty in the number of background events
will be of $0.1 \times 10^7$.
The \Ub{} signal with $0.5 \times 10^7$ event statistics 
(five times larger than background statistical fluctuation)
corresponds to $0.6 \times 10^{-7}$ of the annihilation events
or a square of coupling constant $|f_{_{eU}}|^2 \,=\, 3 \cdot 10^{-9}$.
Such sensitivity will present up to 1500 times improvement
for the mass $m_{_U} = 10$~MeV with the vector coupling compared 
to the upper limit obtained from the measurement of the electron ($g-2$).

The author is grateful to the organizers of the JPOS-2009 conference for their
invitation, K.~de~Jager and J.~LeRose for discussions, and 
P.~Degterenko for help with MC simulation of the experiment. 
This work was supported by the U.S. Department of Energy.
Jefferson Science Association operates the Thomas Jefferson National 
Accelerator Facility for the DOE under contract DE-AC05-060R23177.
\vspace{-0.15in}
\bibliographystyle{aipproc}

\end{document}